\documentclass[twoside,twocolumn,9pt]{article}
\usepackage{extsizes}
\usepackage[super,sort&compress,comma]{natbib} 
\usepackage[version=3]{mhchem}
\usepackage[left=1.5cm, right=1.5cm, top=1.785cm, bottom=2.0cm]{geometry}
\usepackage{balance}
\usepackage{times,mathptmx}
\usepackage{sectsty}
\usepackage{graphicx} 
\usepackage{lastpage}
\usepackage[format=plain,justification=justified,singlelinecheck=false,font={stretch=1.125,small,sf},labelfont=bf,labelsep=space]{caption}
\usepackage{float}
\usepackage{fancyhdr}
\usepackage{fnpos}
\usepackage{bm}
\usepackage{soul}
\usepackage[english]{babel}
\addto{\captionsenglish}{%
	
}
\usepackage{array}
\usepackage{droidsans}
\usepackage{charter}
\usepackage[T1]{fontenc}
\usepackage[usenames,dvipsnames]{xcolor}
\usepackage{setspace}
\usepackage[compact]{titlesec}
\usepackage{hyperref}
\usepackage{epstopdf}%This line makes .eps figures into .pdf - please comment out if not required.

\definecolor{cream}{RGB}{222,217,201}

\begin{document}

\pagestyle{fancy}
\thispagestyle{plain}
\fancypagestyle{plain}{
	
	%%%HEADER%%%
	\fancyhead[C]{\includegraphics[width=18.5cm]{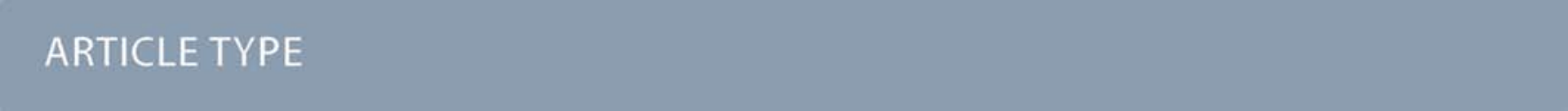}}
	\fancyhead[L]{\hspace{0cm}\vspace{1.5cm}\includegraphics[height=30pt]{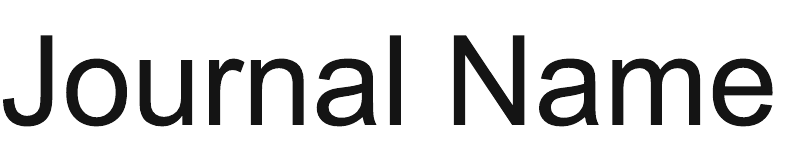}}
	\fancyhead[R]{\hspace{0cm}\vspace{1.7cm}\includegraphics[height=55pt]{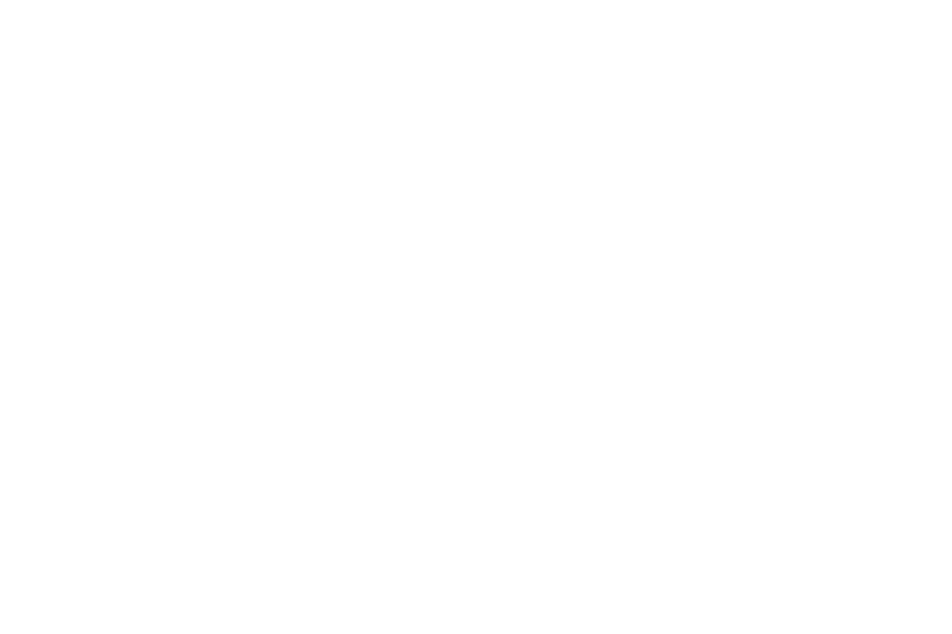}}
	\renewcommand{\headrulewidth}{0pt}
}
%%%END OF HEADER%%%

%%%PAGE SETUP - Please do not change any commands within this section%%%
\makeFNbottom
\makeatletter
\renewcommand\LARGE{\@setfontsize\LARGE{15pt}{17}}
\renewcommand\Large{\@setfontsize\Large{12pt}{14}}
\renewcommand\large{\@setfontsize\large{10pt}{12}}
\renewcommand\footnotesize{\@setfontsize\footnotesize{7pt}{10}}
\makeatother

\renewcommand{\thefootnote}{\fnsymbol{footnote}}
\renewcommand\footnoterule{\vspace*{1pt}% 
	\color{cream}\hrule width 3.5in height 0.4pt \color{black}\vspace*{5pt}} 
\setcounter{secnumdepth}{5}

\makeatletter 
\renewcommand\@biblabel[1]{#1}            
\renewcommand\@makefntext[1]% 
{\noindent\makebox[0pt][r]{\@thefnmark\,}#1}
\makeatother 
\renewcommand{\figurename}{\small{Fig.}~}
\sectionfont{\sffamily\Large}
\subsectionfont{\normalsize}
\subsubsectionfont{\bf}
\setstretch{1.125} %In particular, please do not alter this line.
\setlength{\skip\footins}{0.8cm}
\setlength{\footnotesep}{0.25cm}
\setlength{\jot}{10pt}
\titlespacing*{\section}{0pt}{4pt}{4pt}
\titlespacing*{\subsection}{0pt}{15pt}{1pt}
%%%END OF PAGE SETUP%%%

%%%FOOTER%%%
\fancyfoot{}
\fancyfoot[LO,RE]{\vspace{-7.1pt}\includegraphics[height=9pt]{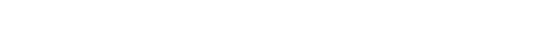}}
\fancyfoot[CO]{\vspace{-7.1pt}\hspace{13.2cm}\includegraphics{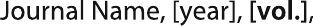}}
\fancyfoot[CE]{\vspace{-7.2pt}\hspace{-14.2cm}\includegraphics{head_foot/RF}}
\fancyfoot[RO]{\footnotesize{\sffamily{1--\pageref{LastPage} ~\textbar  \hspace{2pt}\thepage}}}
\fancyfoot[LE]{\footnotesize{\sffamily{\thepage~\textbar\hspace{3.45cm} 1--\pageref{LastPage}}}}
\fancyhead{}
\renewcommand{\headrulewidth}{0pt} 
\renewcommand{\footrulewidth}{0pt}
\setlength{\arrayrulewidth}{1pt}
\setlength{\columnsep}{6.5mm}
\setlength\bibsep{1pt}
%%%END OF FOOTER%%%

%%%FIGURE SETUP - please do not change any commands within this section%%%
\makeatletter 
\newlength{\figrulesep} 
\setlength{\figrulesep}{0.5\textfloatsep} 

\newcommand{\topfigrule}{\vspace*{-1pt}% 
	\noindent{\color{cream}\rule[-\figrulesep]{\columnwidth}{1.5pt}} }

\newcommand{\botfigrule}{\vspace*{-2pt}% 
	\noindent{\color{cream}\rule[\figrulesep]{\columnwidth}{1.5pt}} }

\newcommand{\dblfigrule}{\vspace*{-1pt}% 
	\noindent{\color{cream}\rule[-\figrulesep]{\textwidth}{1.5pt}} }

\makeatother
%%%END OF FIGURE SETUP%%%

%%%TITLE, AUTHORS AND ABSTRACT%%%
\twocolumn[
\begin{@twocolumnfalse}
	\vspace{3cm}
	\sffamily
	\begin{tabular}{m{4.5cm} p{13.5cm} }
		
		\includegraphics{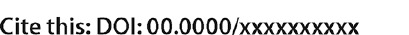} & \noindent\LARGE{\textbf{Chromosome Compaction and Chromatin Stiffness Enhance Diffusive Loop Extrusion by Slip-Link Proteins$^\dag$}} \\%Article title goes here instead of the text "This is the title"
		\vspace{0.3cm} & \vspace{0.3cm} \\

		& \noindent\large{A.~Bonato,$^{\ast}$\textit{$^{a}$} C.~A. Brackley,\textit{$^{a}$} J. Johnson,\textit{$^{a}$} D. Michieletto,\textit{$^{a,b,c}$} and D. Marenduzzo$^{\ddagger}$\textit{$^{a}$}} \\%Author names go here instead of "Full name", etc.
		
		\includegraphics{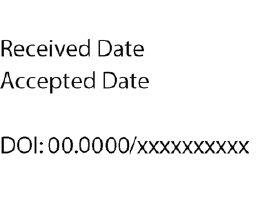} & \noindent\normalsize{We use Brownian dynamics simulations to study the formation of chromatin loops through diffusive sliding of slip-link-like proteins, mimicking the behaviour of cohesin molecules. We recently proposed that diffusive sliding is sufficient to explain the extrusion of chromatin loops of hundreds of kilo-base-pairs (kbp), which may then be stabilised by interactions between cohesin and CTCF proteins. Here we show that the flexibility of the chromatin fibre strongly affects this dynamical process, and find that diffusive loop extrusion is more efficient on stiffer chromatin regions. We also show that the dynamics of loop formation are faster in confined and collapsed chromatin conformations but that this enhancement is counteracted by the increased crowding. We provide a simple theoretical argument explaining why stiffness and collapsed conformations favour diffusive extrusion. In light of the heterogeneous physical and conformational properties of eukaryotic chromatin, we suggest that our results are relevant to understand the looping and organisation of interphase chromosomes \emph{in vivo}. } \\		
	\end{tabular}
\end{@twocolumnfalse} \vspace{0.6cm}
]
%%%END OF TITLE, AUTHORS AND ABSTRACT%%%

%%%FONT SETUP - please do not change any commands within this section
\renewcommand*\rmdefault{bch}\normalfont\upshape
\rmfamily
\section*{}
\vspace{-1cm}

%%%FOOTNOTES%%%

\footnotetext{\textit{$^{a}$SUPA, School of Physics and Astronomy, University of Edinburgh, Peter Guthrie Road, Edinburgh, EH9 3FD, UK}}
\footnotetext{\textit{$^{b}$~MRC Human Genetics Unit, Institute of Genetics and Molecular Medicine, University of Edinburgh, Edinburgh EH4 2XU, UK}}
\footnotetext{\textit{$^{c}$~Department of Mathematical Sciences, University of Bath, North Rd, Bath BA2 7AY, UK}}
\footnotetext{$\ddagger$ Email: davide.marenduzzo@ed.ac.uk}
\footnotetext{$\ast$ Email: A.Bonato@sms.ed.ac.uk}

%Please use \dag to cite the ESI in the main text of the article.
%If you article does not have ESI please remove the the \dag symbol from the title and the footnotetext below.
\footnotetext{\dag~Electronic Supplementary Information (ESI) available: [details of any supplementary information available should be included here]. See DOI: 00.0000/00000000.}
%additional addresses can be cited as above using the lower-case letters, c, d, e... If all authors are from the same address, no letter is required

%\footnotetext{\ddag~Additional footnotes to the title and authors can be included \textit{e.g.}\ `Present address:' or `These authors contributed equally to this work' as above using the symbols: \ddag, \textsection, and \P. Please place the appropriate symbol next to the author's name and include a \texttt{\textbackslash footnotetext} entry in the the correct place in the list.}

%%%END OF FOOTNOTES%%%

%%%MAIN TEXT%%%%

\section*{Introduction}

Chromosome conformation captures (3C) techniques, and their high-throughput variant Hi-C, have provided rich information on the 3-dimensional (3D) organisation of chromosomes in different organisms and cell types~\cite{Dekker2002a,Dixon2012,Rao2014,Giorgetti2014}. In spite of the amount of data produced in recent years, there are still open questions surrounding the biophysical principles that regulate genome organisation \emph{in vivo}. Ultimately, the goal is to establish the relationship between such 3D structure and genome function and gene expression.
 
Hi-C experiments are able to give information on the (often population averaged) spatial organisation of chromosomes in vivo; they have shown that the genomes of a number of organisms are organised into regions which display enriched self-interaction, called ``topologically-associating domains'', or TADs~\cite{Dixon2012}. In mammals, an important class of TADs are those enclosed within a chromatin loop bringing together binding sites of the zinc-finger protein CCCTC binding factor (CTCF)~\cite{Phillips2009,Tang2015,Oti2016}. These contacts are such that they form the base of a loop and establish TAD boundaries. CTCF binding sites often show enrichment for cohesin~\cite{Nasmyth2011,Ocampo-Hafalla2011,Uhlmann2016,Hirano2006}, an SMC protein with a ring-like structure, originally identified for its role in sister chromatid cohesion. Cohesin is thought to be able to bind chromatin by topologically embracing it, and to be able to hold together two segments to stabilize a loop~\cite{Nasmyth2011,Uhlmann2016,Hirano2016}.

\begin{figure}[t!]
	\begin{center}
		\centerline{\includegraphics[width=.95\columnwidth]{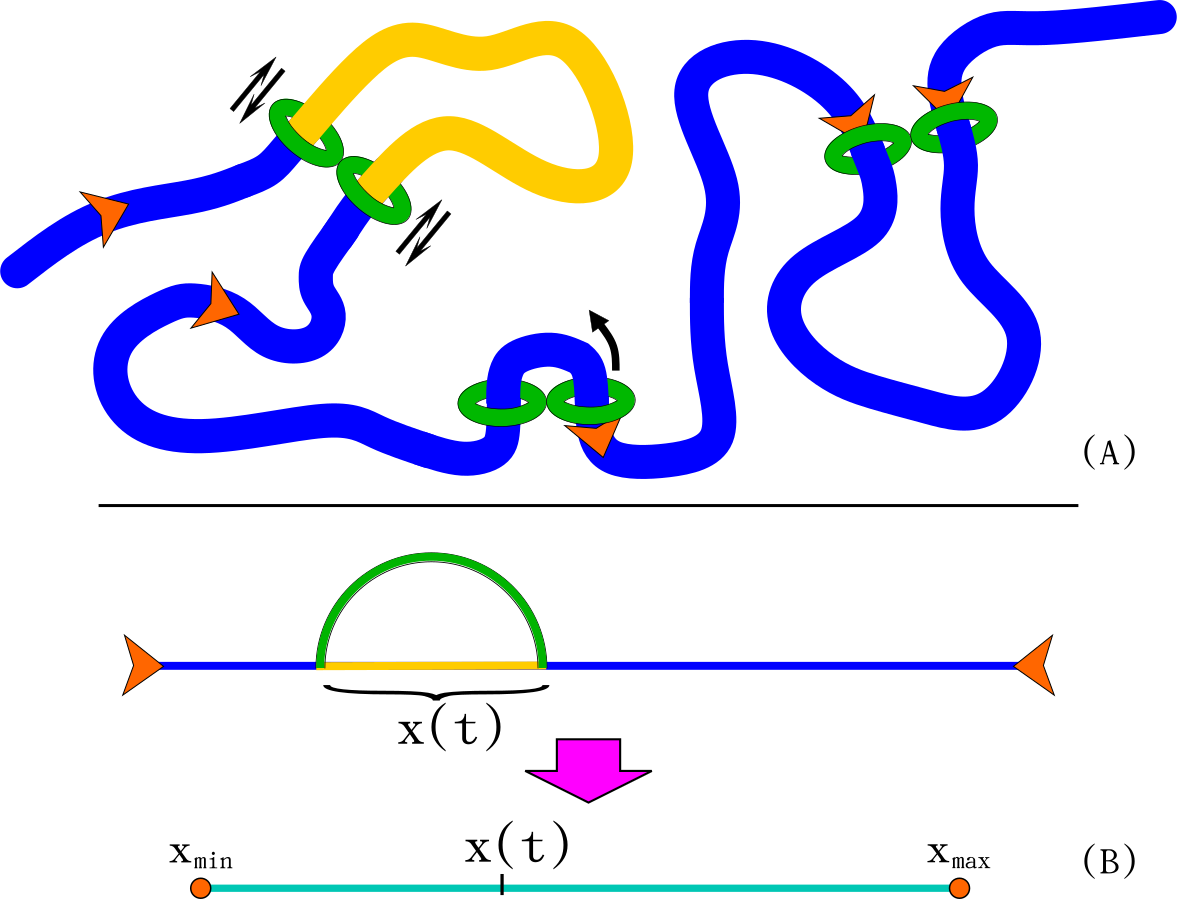}}
	\end{center}
	\caption{(A) Illustration of our model: cohesins are viewed as diffusing handcuffs which subtend loops that can extrude or shrink. When a cohesin meets the convergent side of a CTCF it binds irreversibly.   (B) Diagrammatic representation of the 3D model and mapping to the 1D model: the green arc joins the segments bound by the handcuff, i.e. the base of the loop. In the 1D model, the length of the loop extruded by the cohesin at time $t$ (yellow) is mapped to the position of a random walker moving in the segment $[x_{min}, x_{max}]$. }
	\label{fig1}
\end{figure}

As the binding sequence for CTCF is non-palindromic, it has an orientation along the chromatin. Hi-C experiments revealed that in the vast majority of cases ($>90\%$) the two CTCF binding sites at the base of a loop have a convergent orientation~\cite{Rao2014}. 
This puzzling bias for convergent loops cannot be explained if these binding sites come together through 3D diffusion, but can be reconciled with a ``loop extruding'' mechanism~\cite{Fudenberg2016,Sanborn2015a,Alipour2012,Nasmyth2011}. In this model, cohesin (or another bivalent loop-extruding factor) is able to bind chromatin and actively move along the fibre in such a way that the genomic distance between the segments brought together by the extrusion factor, i.e. the loop length, grows linearly in time (Fig.~\ref{fig1}). If loop extrusion is halted when cohesin meets a CTCF whose binding site is oriented towards it then the convergent bias is naturally explained. 

Another SMC protein, condensin, has been shown to be able to actively extrude loops on DNA \emph{in vitro}~\cite{Ganji2018}, and recent studies showed that cohesin can also actively extrude loops on DNA~\cite{Kim2019,Davidson2019}.  Though they were obtained after histones were depleted, experiments in \emph{Xenopus} egg extracts showed that these two SMC proteins display different extrusion behaviour: both cohesin and condensin extrude actively, the first in a one-sided manner whereas the second symmetrically\cite{Golfier2019}.
	
Despite these advances, the role of active extrusion in a chromatin context \emph{in vivo} remains unclear. \emph{In vitro} evidence suggests that these motors are weak~\cite{Ganji2018}, and so may become detached from chromatin if subject to forces generated by the cellular machines. It is also unclear how an active extruder could maintain its direction over long distances. \emph{In vivo} experiments where ATP is depleted from cells show a loss of loops~\cite{Vian2018a}, but it is unclear whether ATP is required for extrusion, or only for loading and formation of a small initial loop. One might then consider whether an active motor action is strictly necessary for loop extrusion.

We recently showed that it is not~\cite{Brackley2017prl,Brackley2018nucleus}, and proposed an alternative model of diffusive loop extrusion, where cohesin binds to the chromatin fibre and diffuses until it either unbinds or sticks to a bound CTCF protein. As in the active loop extrusion model we assumed that the CTCF-cohesin interaction depends on the relative orientation of CTCF, i.e. cohesin diffuses away if a CTCF is pointing away from it~\footnote{This is slightly different from the active case, where the CTCF must act as a ``one way street''. For diffusive extrusion an incorrectly oriented CTCF acts as a reflecting boundary, whereas a correctly oriented one is a sticky boundary.}. A similar, purely diffusive mechanism for cohesin-driven extrusion was proposed in Ref.~\cite{Yamamoto2017}. These diffusive models are able to generate many of the features seen in Hi-C data, including TADs, grids of loops and stripes~\cite{Vian2018a,Sanborn2015a,Fudenberg2016}. As well as passive and diffusive models, other mechanisms have been proposed to dispense of an explicit motor activity for cohesin. For instance in Ref.~\cite{Racko2018a,Benedetti2014,Benedetti2017,Naughton2013} the authors suggest that supercoiling generated by transcription is sufficient to power the extrusion process. 
While the details of the mechanism \emph{in vivo} remains unclear, it is important to study the properties of all of the possible models.

In Ref.~\cite{Brackley2017prl} it was shown that purely diffusive loop extrusion can lead to the formation of a $100$-kbp convergent CTCF loop within $\sim 20$ min, i.e. within the measured mean cohesin residence time on chromatin~\cite{Stigler2016,Davidson2016,Kanke2016}~\footnote{There is now also evidence that some cohesin molecules can be kept bound for much longer time~\cite{Wutz2019}.}, if the diffusion of cohesin on chromatin is $10$~kbp$^2/$s or more, which appears to be reasonable given recent \textit{in vitro} measurements. For instance, acetylated cohesin was reported to diffuse at $0.1$~$\mu$m$^2$s on reconstituted chromatin~\cite{Kanke2016}, and, assuming a conservative estimate of compaction of $20$~bp/nm on the fibre (which is relevant for an open $10$-nm fibre \textit{in vivo}~\cite{Calladine1997,Alberts2014}), one can infer a diffusion coefficient of $40$~kbp$^2/$s. Recent \emph{in vivo} evidence suggested that a 1Mbp TAD can form within 40 minutes~\cite{Vian2018a}, which would require $\sim 400$ kbp$^2/$s; this is again compatible with our model albeit on either a more compacted chromatin 1kbp/10nm or due to nested loops configurations which enhance the effective diffusion via ratcheting~\cite{Brackley2017prl}.

Irrespective of whether the loop formation by cohesin is an active or diffusive process, it is clear that the biophysical properties of the underlying chromatin substrate can affect the dynamics of cohesin sliding. In this paper we thus aim to further develop a model for cohesin-mediated loop formation by means of Brownian dynamics simulations and focus in particular on the effect of local chromatin stiffness and folding. This is motivated by the fact that the persistence length of chromatin \textit{in vivo} cannot be easily measured; its values estimated experimentally range between $10$ and $200$ nm~\cite{Langowski2006,Sanborn2015a} and are expected to vary across the genome depending on local chromatin fibre structure~\cite{Gilbert2004,Boettiger2016}. In addition to this, local chromatin folding is also variable accross the genome and can attain open and collapsed conformations~\cite{Boettiger2016,Ou2017}.
The contributions from local chromatin stiffness and compaction are expected to affect both, active and diffusive extrusion. 
Here we implement a model for cohesins as diffusing physical handcuffs by means of 3D Brownian dynamics simulations, 1D simulations and theory to study the formation of loops on heterogeneous chromatin fibres. We find that the creation of large loops is favoured on stiff fibres due to enthalpy and that this enhancement holds for collapsed and confined chromatin, such as that found within a eukaryotic nucleus. Even more strikingly, we find that entropic contributions favour long range looping in collapsed conformations but the kinetics of this process is hindered by the increasing microscopic friction. Our work complements previous findings, suggests a route to more physically realistic  (active and diffusive) models for cohesin and condensin extrusion and expands the theoretical framework needed to understand one of the outstanding problems in 3D genome organisation. 

%check figure caption

\section*{Model and methods}
We perform Brownian dynamics (BD) simulations of a chromatin fibre, modelled as a bead-and-spring polymer (with $N=2000$ beads, each of size $\sigma$), where beads are strung together by finite-extension-nonlinear-elastic (FENE) bonds taking the form
\begin{equation}\label{eq:Ufene}
U_{\rm FENE}(r) = \left\{
\begin{array}{lcl}
-0.5kR_0^2 \ln\left(1-(r / R_0)^2\right) & \ r\le R_0 \\ \infty & \
r> R_0 &
\end{array} \right. \, ,
\end{equation}
where $k = 30\epsilon/\sigma^2$ is the spring constant and $R_{0}=1.5\sigma$ is the maximum extension of the bond.
Excluded volume interactions between beads (including consecutive beads along the contour of the chains) are described by the Weeks-Chandler-Andersen (WCA) potential:
\begin{equation}\label{eq:LJ}
U_{\rm WCA}(r) = \left\{
\begin{array}{lr}
4 \epsilon \left[ \left(\frac{\sigma}{r}\right)^{12} - \left(\frac{\sigma}{r}\right)^6 + \frac14 \right] & \, r \le r_c \\
0 & \, r > r_c
\end{array} \right. \, ,
\end{equation}
where $r$ denotes the separation between the bead centers and $r_c=2^{1/6}\sigma$. 
A key role in this work is played by the persistence length, which determines the fibre stiffness. This is introduced through a Kratky-Porod potential, defined in terms of the positions of a triplet of neighbouring beads along the polymer as follows:
\begin{equation}\label{Ubend}
U_{\rm B}(i,i+1,i+2) = \dfrac{k_BT l_p}{\sigma}\left[ 1 - \dfrac{\mathbf{d}_{i,i+1} \cdot \mathbf{d}_{i+1,i+2}}{d_{i,i+1}d_{i+1,i+2}} \right],
\end{equation}
where we denote the position of the centre of the {$i$}-th chromatin bead by {$\mathbf{r}_i$}, the separation vector between beads $i$ and $j$ by $\mathbf{d}_{i,j}=\mathbf{r}_i-\mathbf{r}_j$, and its modulus by $d_{i,j}=|\mathbf{r}_i-\mathbf{r}_j|$.
We used a cubic simulation box and periodic boundary conditions -- the box size is $200\sigma$. In the first part of our work we consider a dilute regime, while in the second part the polymer is confined. 
CTCF binding sites are modelled as stretches of $6$ beads on the polymer which are placed every $100$ beads; we assume that each stretch models a pair of binding sites, and that slip-links strongly bind to the first bead in a stretch facing them, so as to give a directionality to the binding sites and form convergent loops. 

Cohesins are modelled as molecular slip-links formed by two rigid rings, thus our model mimics the case of dimerized cohesin complexes, both sides entrapping a single segment of the fibre~\footnote{We expect that, with some modifications to the model, similar results would be observed for a single ring entrapping two strands.}. Each of these rings is composed of $12$ beads, arranged in a square (with side $4\sigma$), with an additional phantom sphere at the centre which interacts only with beads on the chromatin fibre modelling CTCF binding sites. The two rings are held together by four FENE bonds, and they are kept in an open ``handcuff'' arrangement via two sufficiently strong bending interactions (the potential has the same functional form as in Eq.~(\ref{Ubend})). The CTCF-cohesin interaction is modelled via a Morse potential between the first bead in a CTCF stretch and the phantom bead in the middle of the slip-link rings: 
\begin{equation}\label{eq:Morse}
U_{\rm Morse}(r) = \left\{
\begin{array}{lr}
\epsilon \left[ e^{-2\alpha r}-2e^{-\alpha r} \right] - \epsilon \left[ e^{-2\alpha r_c}-2e^{-\alpha r_c} \right]& \, r \le r_c \\
0 & \, r > r_c
\end{array} \right. \, .
\end{equation}
We set $\epsilon = 10.0$ $k_BT$, the range $r_c = 1.2\sigma$ and $\alpha = 3\sigma^{-1}$. These values ensure the interaction is strong enough that once cohesin meets a CTCF, they stay bound for the rest of the simulation (i.e. it is our absorbing state).

The motion of the centre of mass of the slip-links, as well as the diffusive motion of polymer beads, are described by a Langevin equation
\begin{equation}
m \dfrac{d^2 \bm{r}_i}{dt^2} = -\zeta \dfrac{d \bm{r}_i}{dt} - \nabla U_i + \sqrt{2 k_BT \zeta} \bm{f}_i
\end{equation}
where $U$ is the total potential experienced by a bead or a cohesin ring, $\zeta$ is the friction on each bead and the components of $\bm{f}$ are independent Gaussian random variables with zero mean and unit variance. The factor $\sqrt{2 k_BT\zeta}$ ensures the system satisfies the fluctuation-dissipation theorem with temperature $T$. A similar rotational equation determines the orientation of the rings. We use LAMMPs molecular dynamics software, which evolves the equations of motion using a velocity-Verlet algorithm~\cite{Plimpton1995}.

\begin{figure}
	\begin{center}
		\centerline{\includegraphics[width=.95\columnwidth]{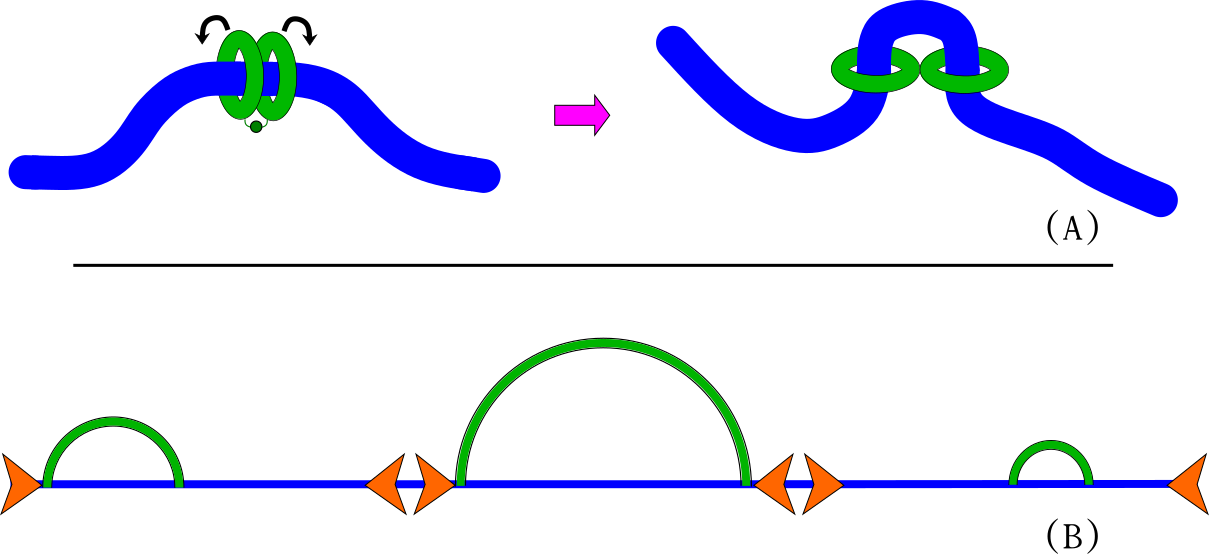}}
	\end{center}
	\caption{(A) This illustration shows how slip-links are loaded. First as a folded pair of rigid rings and then straighten up by a bending potential. (B) In our set up we simulate chromatin fibres divided in adjacent sections, each bounded by a pair of convergent CTCF binding sites. One and only one cohesin is loaded per each section.}
	\label{fig:fig2}
\end{figure}

In our simulations cohesin is initially positioned in a folded handcuff arrangement such that each ring encircles an adjacent polymer bead then, the bending interaction between the two rings is turned on thus opening the handcuff and bending the polymer into a small loop (Fig.~\ref{fig:fig2}). After this step, the slip-link is free to diffuse whilst remaining topologically linked to the chromatin fiber, so that its associated loop may grow or shrink. This seemingly complicated set up is not an attempt to realistically model cohesin handcuff loading, but is necessary to correctly load our model cohesin and avoid numerical singularities due to beads overlapping. We expect the extrusion dynamics to be realistic after loading. We also do not include continual unloading and loading of cohesin, and only have one cohesin handcuff per polymer segment. In this way we study simple extrusion without competition or cooperative effects. Such collective effects have been investigated in our previous work~\cite{Brackley2017prl}.

\begin{figure}[t]
	\begin{center}
		\centerline{\includegraphics[width=\columnwidth]{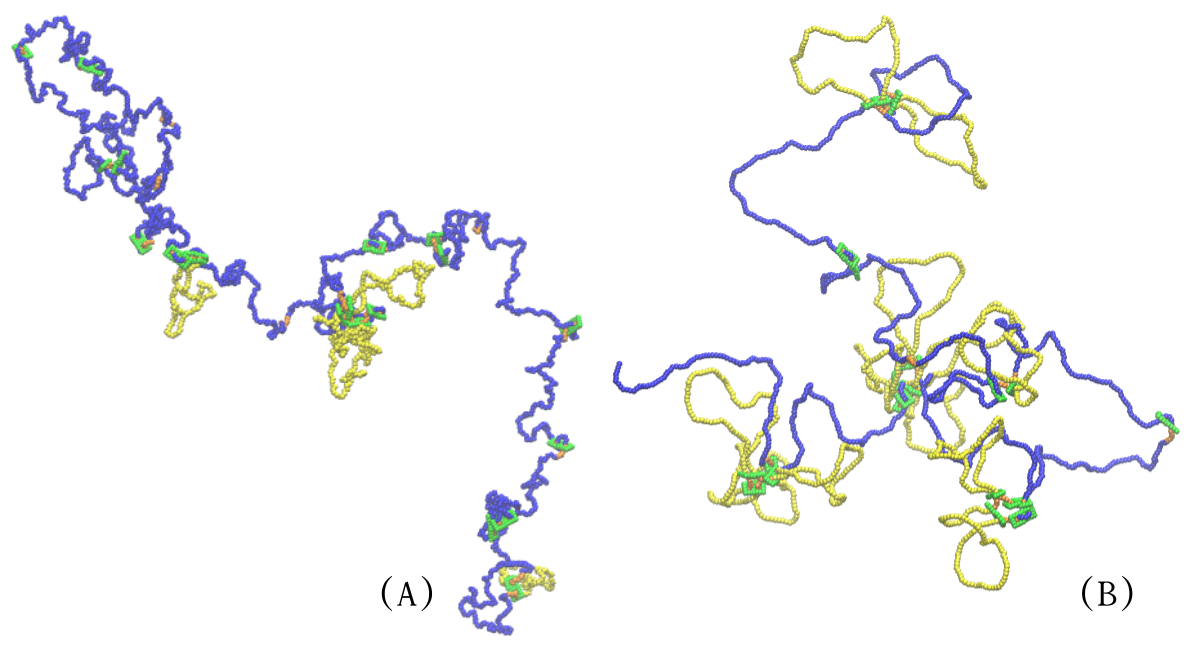}}
	\end{center}
	\caption{Snapshots from 3D simulations of (A) a flexible (persistence length $2\sigma$) and (B) a semi-flexible (persistence length $10\sigma$) chromatin fibre. Blue and orange beads correspond to standard chromatin beads and CTCF binding sites respectively, whereas beads making up slip-links are depicted in green. Complete chromatin loops (between two neighbouring CTCF binding sites) are highlighted in yellow. Notice the 3D clustering of CTCFs driven by looping of contiguous domains.}
	\label{fig3}
\end{figure}
We model confinement by enforcing that the polymers must remain within a sphere of radius $R$. Beads that attempt to escape this sphere are subjected to a harmonic restoring force with spring constant $\kappa=\epsilon$. Initial configurations are constructed by progressively confining a fibre which is first equilibrated in a dilute regime and loaded with cohesins stuck in a fixed position along the fibre. This is achieved by changing the diameter $300$ times, each followed by a quick equilibration run of $100$~timesteps, starting from a sphere of radius $300\sigma+R$. Once the desired radius ($R$) is reached, a longer equilibration run of $1.5\times10^6$~timesteps is performed, at the end of which the cohesins are freed to move from their initial position.

The mapping from simulation to physical units can be made as follows. Energies are measured in units of $k_B T$. To map length scales from simulation to physical units, we set the diameter, $\sigma$, of each bead to, for instance, $\sim 15$~nm~$\simeq 1$~kbp (assuming a chromatin fibre with compaction intermediate between a $10$~nm and a $30$~nm fibre; all of our results would remain qualitatively unchanged with a different mapping). 
The $l_p$ values we consider are between $2\sigma$ and $10\sigma$ (see snapshots in Fig.~\ref{fig3}), hence they correspond to $\sim 30-150$ nm. These values are within the range expected for chromatin~\cite{Langowski2006}.
To map time units, we need to estimate the typical diffusive timescale (over which a bead diffuses a distance comparable to its own size), or Brownian time, which equals $\tau_B\equiv \sigma^2/D$. One way to do this is to require that the mean square displacement of a polymer bead matches that of a chromatin segment measured {\it in vivo} in Ref.~\cite{Hajjoul2013}. This is similar to the scheme used in Refs.~\cite{Brackley2017prl,Rosa2008,Michieletto2016prx}, and it should be noted that, in this way, we match the effective {\it in vivo} viscosity, taking into account any macromolecular crowding within the nucleoplasm. 
Simulations were run for up to $10^6$~$\tau_B$, and integration was performed with a step size of $0.01$~$\tau_B$.

\section*{Results}
\begin{figure}[t]
	\begin{center}
		\centerline{\includegraphics[width=\columnwidth]{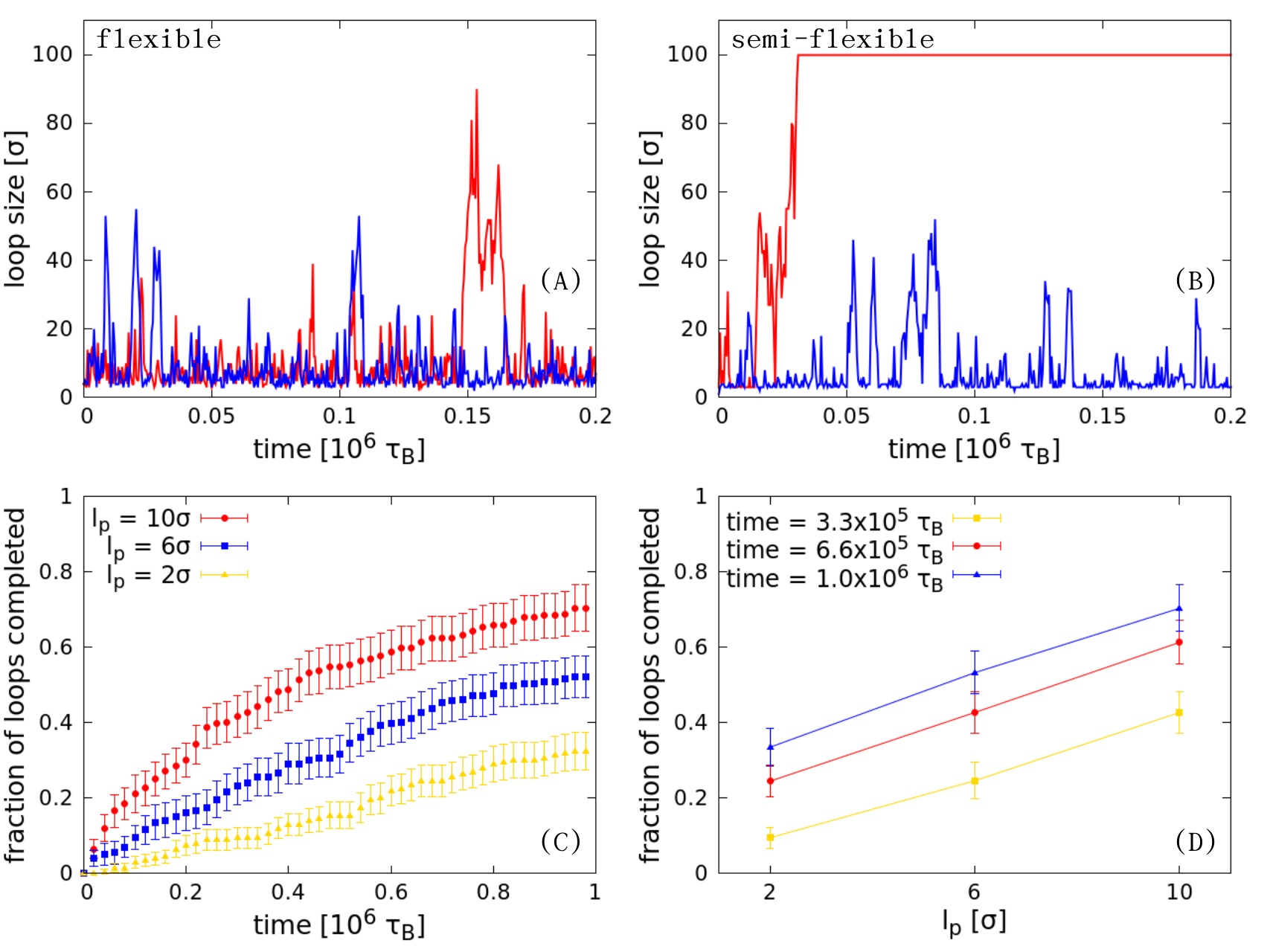}}
	\end{center}
	\caption{Results from 3D simulations of a chromatin fibre split into sections by CTCF sites, with a single slip-link per section. (A) Example of loop size as a function of time for two selected slip-links (blue and red curves) on a (A) flexible ($l_p=2\sigma$) and (B) stiff ($l_p=10\sigma$) chromatin fibre. (C) Fraction of completed (i.e. binding both CTCFs) loops as a function of time, for $l_p=2\sigma$, $l_p=6\sigma$, and $l_p=10\sigma$, and (D) as a function of persistence length at given time.}
	\label{fig4}
\end{figure}

\begin{figure}[t]
	\begin{center}
		\centerline{\includegraphics[width=\columnwidth]{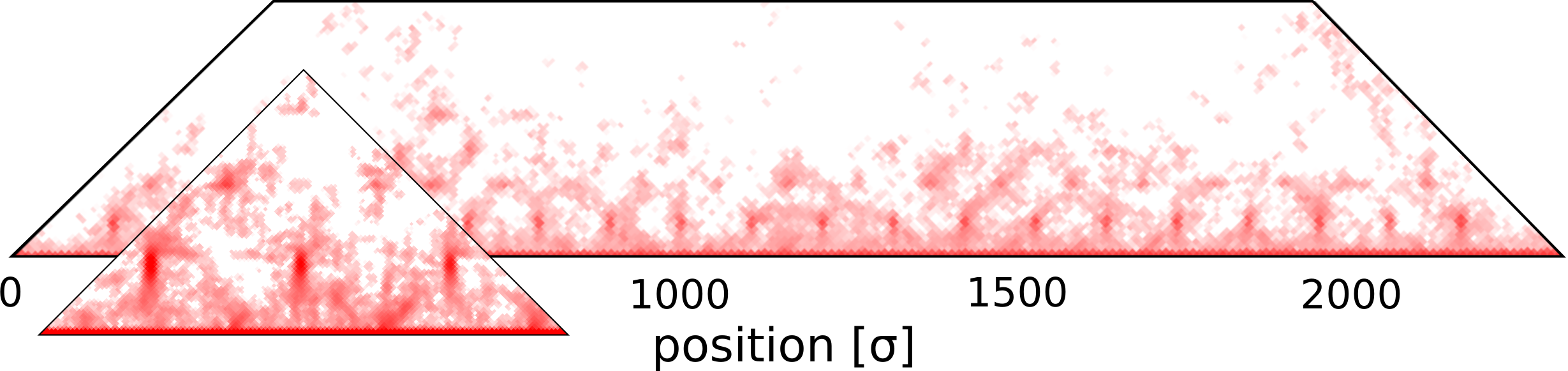}}
	\end{center}
	\caption[]{Contact map obtained from 3D simulations of semi-flexible ($l_p=10\sigma$) fibres in the dilute regime, with zoom on the region $[75\sigma,450\sigma]$  (time $5\times10^5\tau_B$).\footnotemark}
	\label{figCMap}
\end{figure}
\footnotetext{Since in our simulations cohesin dimers bind to CTCF irreversibly, and since we do not model unbinding and rebinding, these results should not be expected to exactly reproduce experimental Hi-C maps.}

\subsection*{Diffusive loop extrusion is more efficient on stiffer chromatin fibres}

We first use BD simulations to study the diffusive sliding of cohesin-like slip-links on a chromatin fibre of variable stiffness and contour length $2.326$ Mbp (corresponding to $2326$ beads including $2200$ standard chromatin beads and $126$ CTCF beads). The model chromatin is split up into 20 sections of $\sim 100$ kbp, each of which is flanked by a convergent pair of CTCF binding sites (these account for the $3$ terminal beads on each end, so that each section is comprised of $100$ chromatin beads and $6$ CTCF beads).  We add $20$ slip-links, one in each section (see Fig.~\ref{fig3}); their initial positions within each section were chosen randomly with a uniform probability.

Figure~\ref{fig4} shows time-series of the size of the loop formed by a cohesin slip-link within a flexible (persistence length $2 \sigma$, Fig.~\ref{fig4}A) and a stiff (persistence length $10 \sigma$, Fig.~\ref{fig4}B) chromatin fibre. The trajectories show that diffusive sliding can create large loops. In particular, such trajectories are unlike those of standard random walks, but are instead characterised by many short excursions and a few larger ones, some of which can lead to successful CTCF loop formation (see Figs.~\ref{fig4}B~and~\ref{figCMap}). As we shall see, this is because the entropic cost of looping acts to limit loop size.

Inspection of the trajectories also suggests that diffusive loop extrusion is more efficient on the stiffer chromatin substrate. To show this more quantitatively, we measured the fraction of completed loops (i.e. reaching the CTCF bounding the region) as a function of time and found that the rate of full loop formation can be up to 8-times larger on stiffer fibres (Fig.~\ref{fig4}C-D). The extent of this effect is perhaps surprising, given there is only a factor of $5$ difference between the stiff and the flexible fibres in our simulations. 

\subsection*{A simple 1D model explains the effect of flexibility on diffusive extrusion}

To understand why chromatin flexibility affects slip-link diffusivity, we analyse a simple 1D model of a random walker (slip-link) loaded at a position on the fibre, and diffusing in an effective potential modelling the entropic and enthalpic ``cost'' associated with looping of a semi-flexible polymer. The position of the random walker at time $t$ represents the instantaneous size of a slip-link loop (i.e., the separation between the two sides of the slip-link). A suitable effective potential (defined up to an additive constant), $V$, is the following~\cite{Ringrose1999,Brackley2017prl},
\begin{equation}\label{1Dpotential}
\frac{V(l)}{k_BT} = \frac{8 l_p}{l^2} + c\log(l),
\end{equation}
where $l$ is the loop size, or position of the random walker, $l_p$ is the persistence length, and $c$ is a universal exponent describing the entropic cost of looping (for phantom polymers without excluded volume, $c=3/2$ in 3D). This functional form captures the competition between the bending energy ``cost'', which decreases monotonically with loop size $l$, and the entropic ``cost'', which increases with $l$. For an ideal flexible polymer, the minimum of the potential will therefore be at $0$. In practice, though, this case is of limited interest as self-avoidance alone is sufficient to create a non-zero effective bending rigidity.

\begin{figure}
\begin{center}
\centerline{\includegraphics[width=\columnwidth]{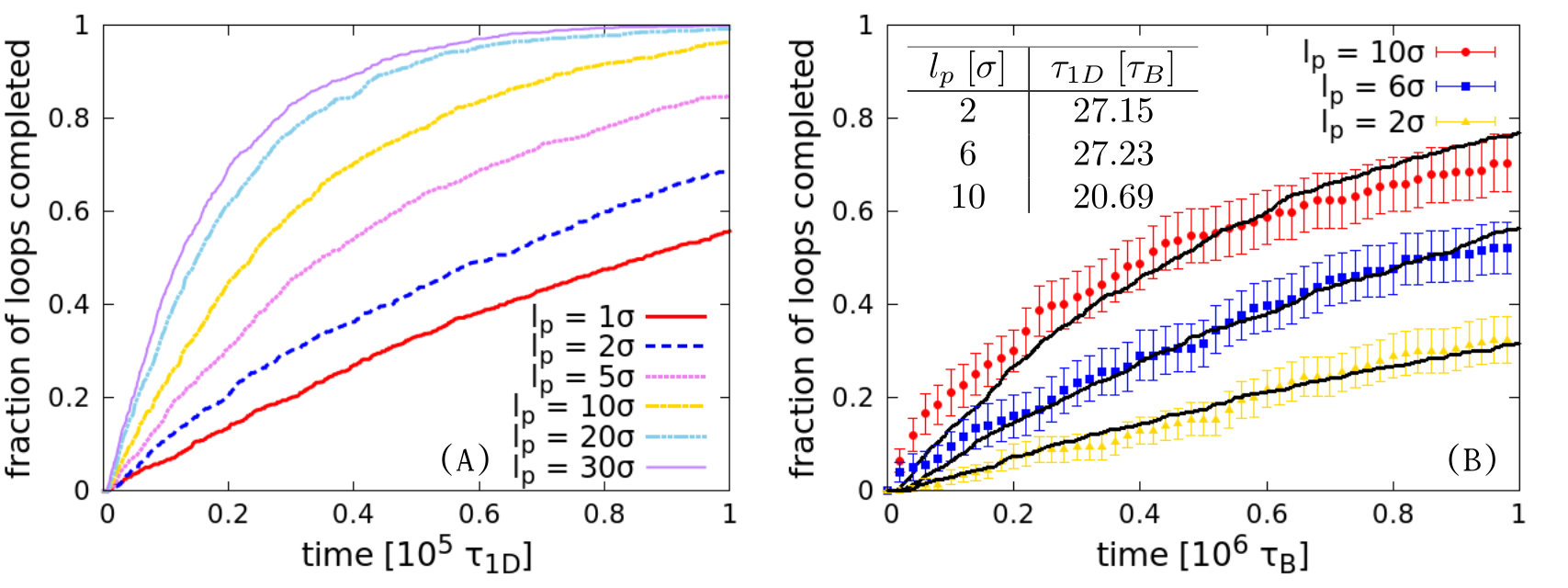}}
\end{center}
\caption{(A) Results from simulations of the 1D model. Fraction of $100$~kbp loops formed for different values of $l_p$, as predicted via numerical evolution of a random walk in 1D model in the potential defined in Eq.~(\ref{1Dpotential}), as a function of time. (B) Comparison between results from 1D and 3D simulations. Solid black lines are fit to the fraction of loops completed as obtained in the 3D simulations with the numerical predictions of the 1D model. The table shows the mapping of $\tau_{1D}$ to $\tau_{B}$ units as obtained via the fit (right column) for each value of the persistence length. }
\label{fig5} 
\end{figure}

In the 1D model, the random walker moves within a domain of size $L$, representing a chromatin section flanked by convergent CTCF sites as in our 3D simulations. We simulate this simple 1D problem by sampling Ito increments for a walker starting at $l=1\sigma$, imposing reflective and absorbing boundary condition at $l=1\sigma$ and $l=100\sigma$ respectively. The diffusion coefficient of the walker is $D=1$~$\sigma^2/\tau{1D}$ and integration was performed with a step size of $0.01$~$\tau_{1D}$. We can find the probability that a CTCF loop has formed (i.e. the walker first reaches position $l=L$) as a function of time, as in our 3D simulations. The associated curve is plotted in Figure~\ref{fig5} for different values of the bending rigidity. As the chromatin stiffness contribution favours loop enlargement when the loop is small, we find that this 1D model qualitatively reproduces the bias in favour of larger loops for stiffer fibres which is observed in the 3D simulations (Fig.~\ref{fig5}).

To quantitatively compare the results of 1D and 3D simulations we to find a mapping between the time units for each model, i.e. we match the curves obtained by 1D simulations to the correspondent ones from 3D BD simulations for each value of the persistence length (Fig.~\ref{fig5}B).
The table of Figure~\ref{fig5}B shows that $\tau_{1D}$ is close to $28$~$\tau_B$, which corresponds to the baseline cohesin diffusivity  $D=\frac{1}{28}$~$\sigma^2/\tau_B$; we remark that $28$ is the mass, expressed in computational units, of a cohesin dimer as modelled in our simulations. 

It should be noted that this model neglects the fact that the loop is made by two diffusing rings. When one of the two becomes bound to a CTCF, the diffusion constant of the subtended loop would halve. This effect is not incorporated into the 1D model and it does not qualitatively affect our results. In addition, as we show below, the complexes diffuse on average $100$ kbp in $10^5$~$\tau_B$ (Fig.~\ref{figMSD}) and thus, after this initial transient, it is safe to assume that at least one of the two rings is bound to a CTCF. This implies that the diffusion coefficient of the one remaining searching ring is also that of the subtended loop which we implement in the 1D model. 

\begin{figure}[t]
	\begin{center}
		\centerline{\includegraphics[width=\columnwidth]{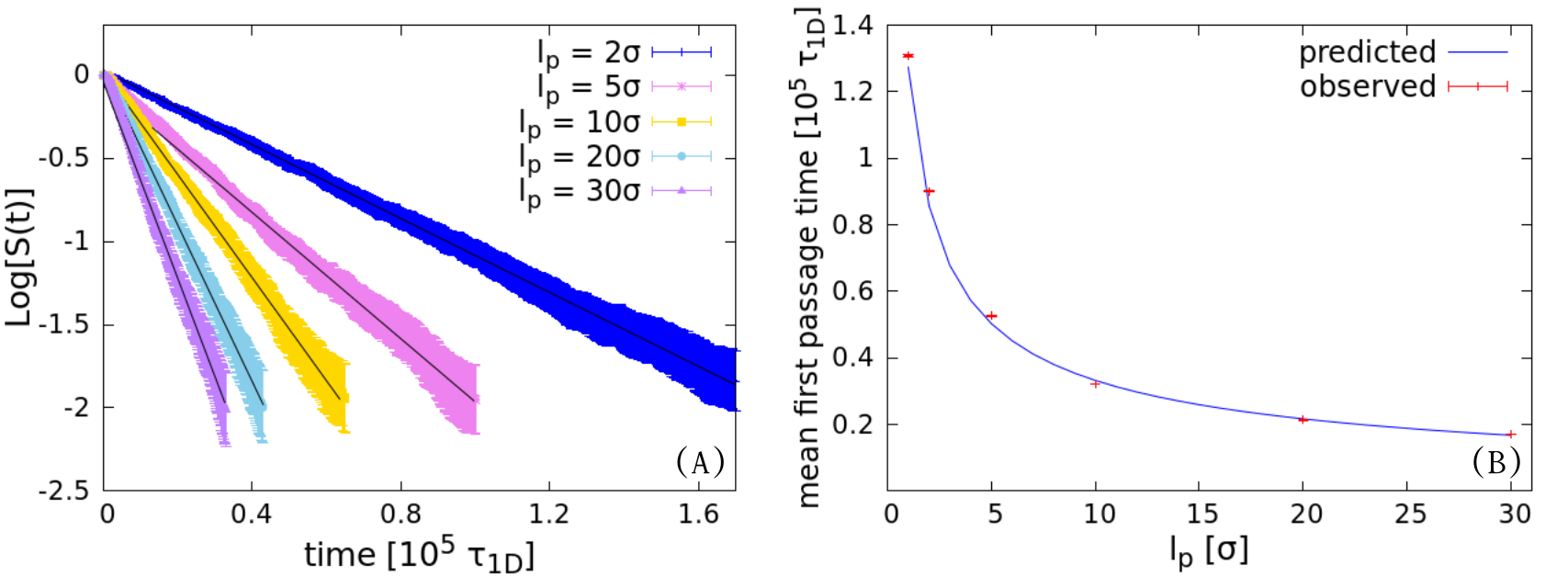}}
	\end{center}
	\caption{Survival probability distributions and mean first passage times as predicted by the 1D model. (A) Logarithm of the survival probability as a function of time for different values of the persistence length. As predicted by Eq.~\eqref{survP}, the decay in time of the survival probability is well fitted by an exponential curve. (B) Comparison between the mean first passage times predicted by Eq.~\eqref{MFPT} and the values obtained from the simulations.}
	\label{fig1Surv} 
\end{figure}

\subsection*{Survival probabilities and extrusion rates}
We now aim to compare the predictions of the 1D and 3D simulations by describing how the number of completed loops (Figs.~{\ref{fig4}}C-{\ref{fig5}}A) evolves with time.
To this end, let us call $T_L(x)$ the mean first passage time to reach point $L$ of a random walker moving in a potential $V$ starting from the initial position $x$.
If $l_{SL}$ is the initial length of the loop enclosed by a handcuff loaded on a segment of length $L$, then $T_L(l_{SL})$ represents the mean time required for a cohesin to completely extrude that segment.
It can be shown (e.g.~\cite{Hanggi1990}) that $T_L(x)$ obeys the following equation (derived from a backwards Fokker-Plank equation)
\begin{equation}\label{bFokkerPlank}
-1=-\left[\frac{1}{\gamma}\frac{dV(x)}{dx}\right]\frac{dT_L(x)}{dx}+D\frac{d^2T_L(x)}{dx^2},
\end{equation}
where $\gamma$ and $D$ are the drag and diffusion coefficients describing the motion of the random walker. By imposing $T_L(L)=0$ and the reflecting boundary condition $\frac{dT_L(l_{SL})}{dx}=0$, the solution of Eq.~\eqref{bFokkerPlank} reads
\begin{equation}\label{Sol}
T_L(x) = \frac{1}{D}\int_{x}^{L}e^{V(y)/k_BT}dy\int_{l_{SL}}^{y}e^{-V(z)/k_BT}dz \, ,
\end{equation}	
and by expressing $V$ as per Eq.~\eqref{1Dpotential} one finds
\begin{equation}\label{MFPT}
T_L(l_{SL}) = \frac{1}{D}\int_{l_{SL}}^{L}e^{8l_P/y^2}y^cdy\int_{l_{SL}}^{y}e^{-8l_P/z^2}z^{-c}dz \, .
\end{equation}

\begin{figure}
	\begin{center}
		\centerline{\includegraphics[width=0.65\columnwidth]{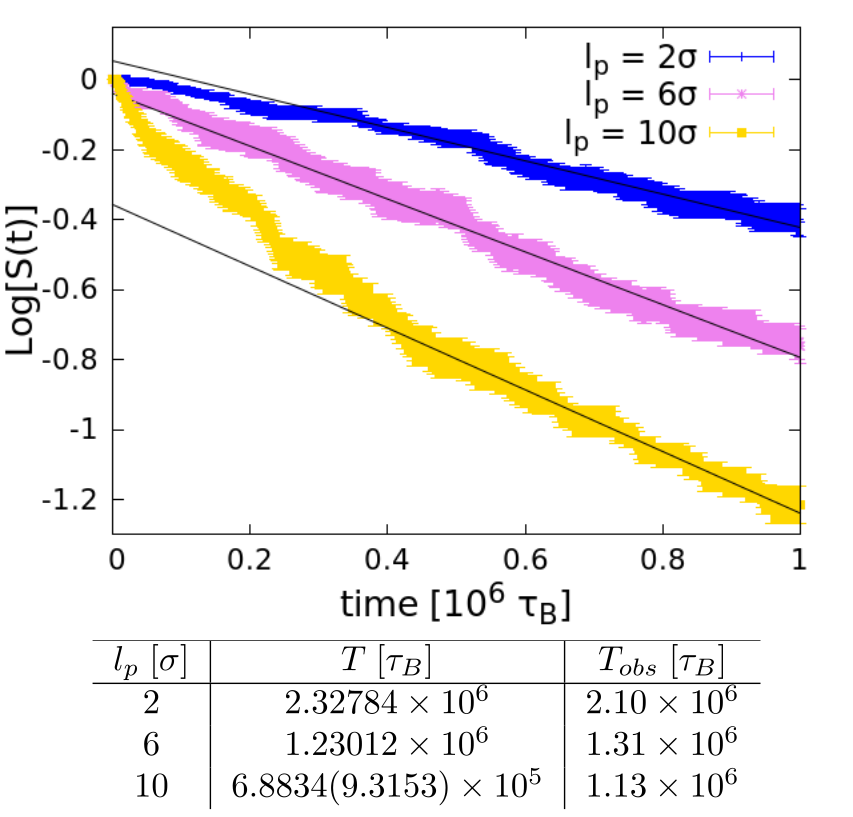}}
	\end{center}
	\caption{Survival probability distributions and mean first passage times in the 3D simulations. Logarithm of the survival probability as a function of time for different values of the persistence length as measured from 3D simulations. The black lines were obtained by fitting the curves starting from time $t=3\times10^5\tau_B$ for $l_p=2,6\sigma$ and $t=5\times10^5\tau_B$ for $l_p=10\sigma$. The second column of the table shows the mean first passage times as predicted by the 1D model, converted to $\tau_B$ units by using the mapping of Figure~\ref{fig5}B; the second value for $l_p=10\sigma$ uses $\tau_{1D}=28$~$\tau_B$ instead. The third column collects the mean first passage times as obtained by the slope of the fitting lines.}
	\label{fig2Surv} 
\end{figure}
Provided that the dynamics over $L$ can be well described by a rate -- i.e. there are no other relevant processes affecting the hopping timescales -- 
the survival probability $S_{L,l_{SL}}(t)$ can be written as~\cite{Hanggi1990}
\begin{equation}\label{survP}
S_{L,l_{SL}}(t) \approx e^{-t/T_L(l_{SL})},
\end{equation}
We numerically solve Eq.~\eqref{MFPT}  and use it to determine the survival probability $S_{L,l_{SL}}(t)$, the probability that at time $t$ the random walker (starting in $l_{SL}$ at time $t=0$) has yet to reach $L$ -- or in other words 1 minus the fraction of completed loops. Figure~\ref{fig1Surv} shows that, after an initial transient, the survival probabilities obtained by simulations of the 1D model are perfectly described by the predictions of Eqs.~\eqref{MFPT} and \eqref{survP}. 

The survival probabilities from the 3D simulations seem to suggest they also decay exponentially in time for large enough fraction of loops completed (Fig.~\ref{fig2Surv}) and the mean first passage times are similar to the correspondent values predicted by the 1D model.

Note that the concordance between the predictions of the 1D model and the BD simulations suggests that the behaviour of the isolated segments is not affected by the presence of the other segments.

\subsection*{Confinement and Collapsed Conformations Enhance Diffusive Extrusion}

\begin{figure}[t]
	\begin{center}
		\centerline{\includegraphics[width=0.95\columnwidth]{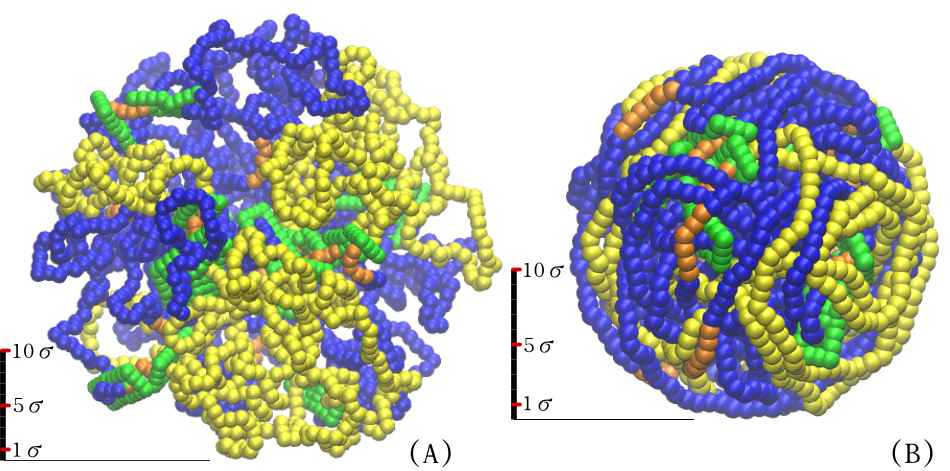}}
	\end{center}
	\caption{Snapshots from 3D simulations of (A) a flexible (persistence length $2\sigma$) chromatin fibre confined in a sphere of radius $20\sigma$ and (B) a semi-flexible (persistence length $10\sigma$) fibre confined in a sphere of radius $12\sigma$. Colours are used as described in Fig.~\ref{fig3}. Scale bars are reported to qualitatively show the strength of confinement in each case.  }
	\label{fig6}
\end{figure}	

An important feature of chromatin {\it in vivo} is that it is under substantial confinement~\cite{Kang2015}. For instance, a human lymphocyte nucleus is about 7 $\mu$m in diameter: if the chromatin fibre is organised into $1$ kbp beads of $15$ nm size (as in our simulations), then its volume fraction in a diploid cell (containing $\sim 6$ Gbp of DNA) is about $5\%$. As DNA packaging (in bp/nm) and nuclear volume are variable, this estimate is only approximate -- using different values for these two quantities within a physiologically relevant range gives volume fractions in the $1-10\%$ range. Additionally, the eukaryotic nucleus is also densely populated by proteins, RNA and nuclear bodies such as the nucleolus. 

It is therefore of interest to ask whether diffusive loop extrusion is a viable mechanism to form convergent CTCF loops for confined polymers, which are in the semidilute or concentrated regime in polymer physics parlance. More generally, it is of interest to study whether differences in polymer conformation, which can be due to confinement but also to \emph{local} compaction due to, e.g., interaction with proteins which bind chromatin with specific histone modifications ~\cite{Brackley2016nar,Michieletto2016prx}, can \emph{locally} affect the kinetics of diffusive loop extrusion. Thus, in this section we focus on fixed flexibility and consider different levels of polymer confinement.

\begin{figure}[t]
	\begin{center}
		\centerline{\includegraphics[width=\columnwidth]{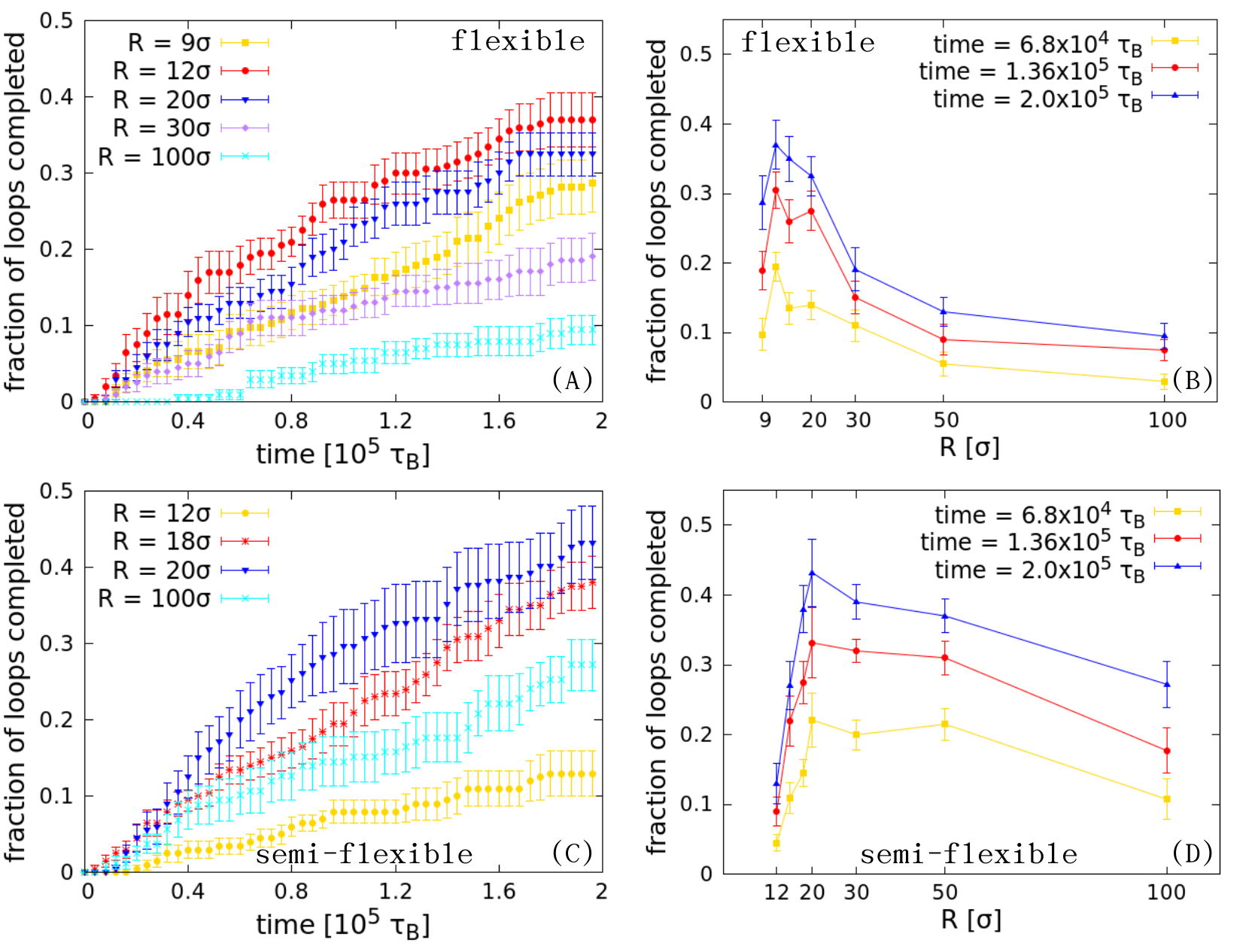}}
	\end{center}
	\caption{Results from 3D simulations of chromatin fibres split into sections by CTCF sites, with a single slip-link per section and confined in spheres of different radius. Fraction of completed loops (A)(C) as a function of time, and (B)(D) confinement radius at given time, for (A)(B) flexible ($l_p = 2\sigma$) fibres and (C)(D) semi-flexible ($l_p = 10\sigma$) fibres.}
	\label{fig7}
\end{figure}

\begin{figure}
	\begin{center}
		\centerline{\includegraphics[width=\columnwidth]{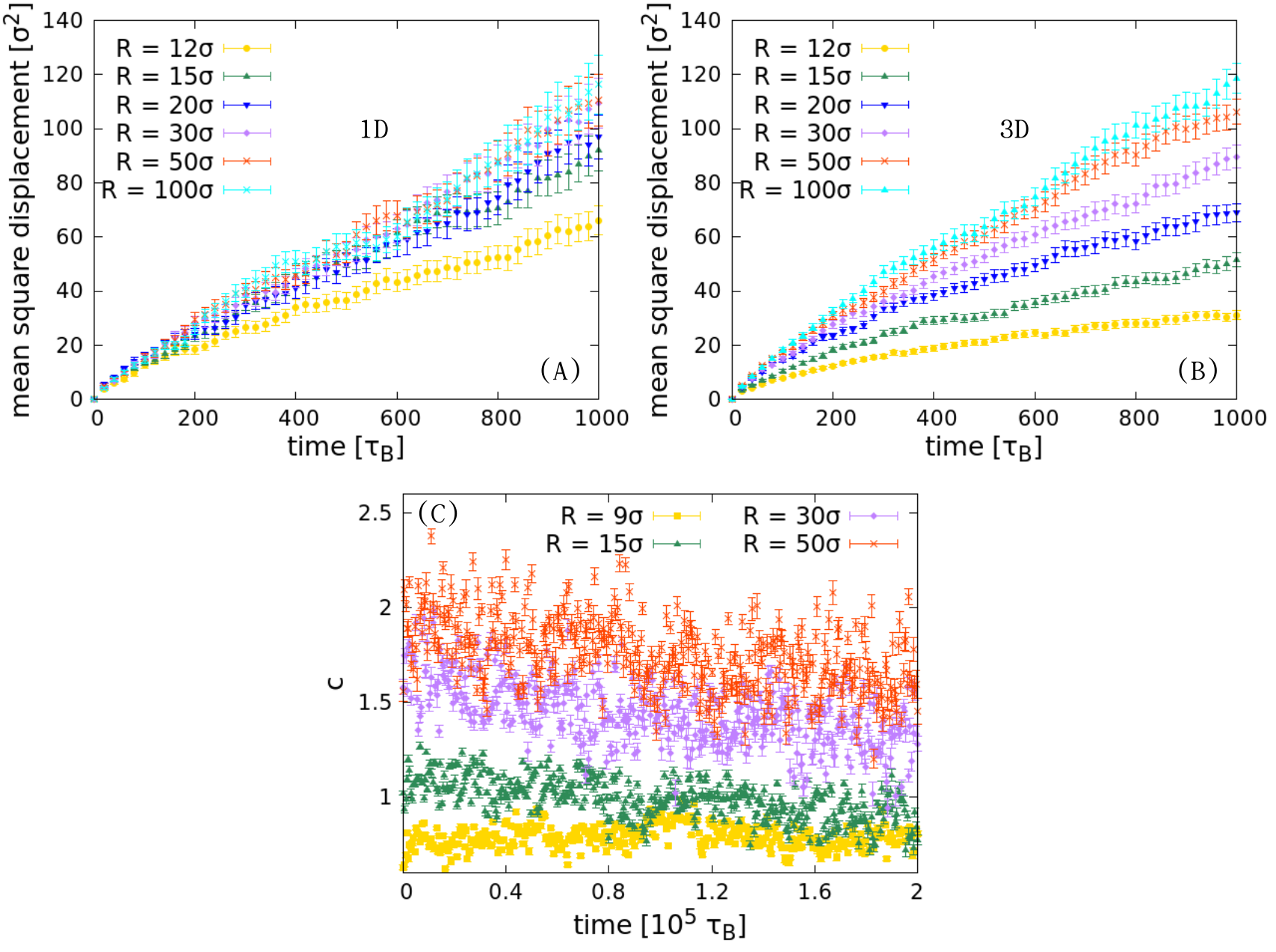}}
	\end{center}
	\caption{Mean square displacement (A) on the fibre and (B) in the 3D space of one of the two rings of a cohesin diffusing along semi-flexible ($l_p=10\sigma$) chromatin fibres confined within spheres of different radius. This was obtained by tracking the position along the fibre -- i.e. the nearest chromatin bead -- and in space of the central phantom beads of the rings. (C) Plot of the contact exponent $c$ as obtained from BD simulations of confined flexible fibres as a function of time.}
	\label{figMSD}
\end{figure}

Confinement might be expected to slow down most dynamical processes associated with diffusive motion, due to crowding effects. Interestingly, we instead observe that, for a certain range of $R$, confinement can also have the opposite effect. As we decrease the confinement radius, diffusive loop extrusion creates CTCF loops more quickly (Fig.~\ref{fig7}). This effect persists down to $R=20$ $\sigma$, corresponding to a volume fraction of $3\%$, and the increase in looping efficiency -- probability of full 100 kbp loop formation at a given time -- due to geometric confinement is about 2-fold. For smaller values of $R$, crowding effects take over, microscopic friction becomes larger and diffusive extrusion progressively slower (Fig.~\ref{figMSD}).

\begin{figure}[t]
	\begin{center}
		\centerline{\includegraphics[width=\columnwidth]{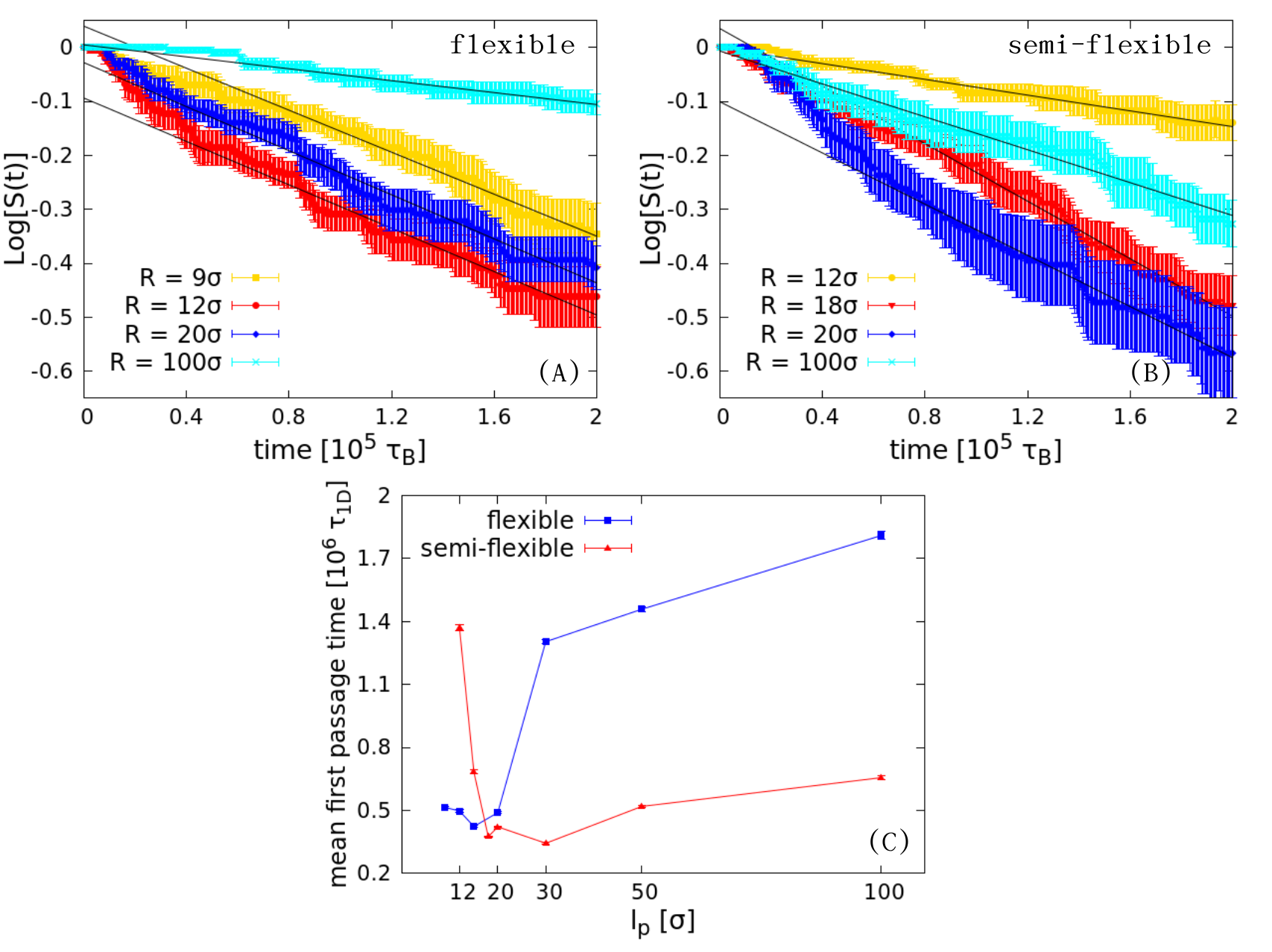}}
	\end{center}
	\caption{Survival probability distributions and mean first passage times in the 3D simulations of confined fibres. Logarithm of the survival probability as a function of time for different values of the confinement radius for (A) flexible ($l_p = 2\sigma$) and (B) semi-flexible ($l_p = 10\sigma$) fibres. The black lines are the result of the linear interpolation of the curves starting from $\bar{t}=6\times10^4\tau_B$. The table contains the measured mean first passage times.}
	\label{fig3Surv} 
\end{figure}

Figure~\ref{fig3Surv} suggests the survival probability of cohesins diffusing along confined fibres still decays exponentially in time after an initial transient, and the non-monotonic behaviour highlighted in Fig.~\ref{fig7} is once again remarked by how the measured values of the mean first passage times change as the confinement radius decrease.

\begin{figure}[t]
	\begin{center}
		\centerline{\includegraphics[width=\columnwidth]{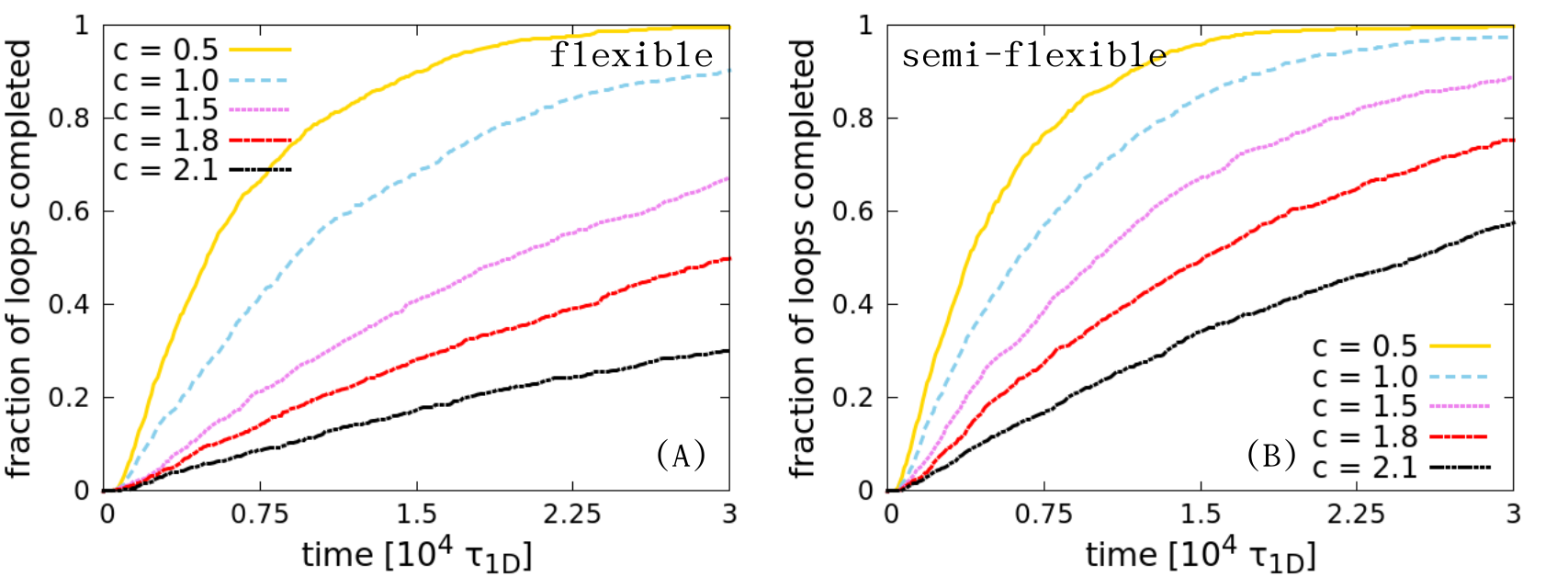}}
	\end{center}
	\caption{Results from simulations of the 1D model. Fraction of $100$~kbp loops formed for different values of $c$, as predicted via numerical evolution of a Random Walk in 1D within a potential defined in Eq.~(\ref{1Dpotential}), as a function of time.}
	\label{fig8}
\end{figure}

To understand the mechanism underlying the increase in efficiency of diffusing loop extrusion, we again turn to the simplified 1D model (using the potential defined in Eq.~\eqref{1Dpotential}). As the polymer is confined, the probability of loop formation is also affected. Above, we added confinement by decreasing the radius of an initially swollen self-avoiding polymer: as the rate at which the final confinement is reached was not adiabatically slow, we expect that the resulting conformation is that of a ``fractal globule''~\cite{Grosberg1988,Grosberg1993,Lieberman-Aiden2009}. In this case, the probability of forming short enough loops would be associated with an exponent $c=1$. Even for an equilibrium globule conformation, the $c$ exponent would vary between $1.5$ and $0$ (at small and large loop length respectively~\cite{Tamm2005}), hence it would still lie below the value of the self-avoiding walk ($c\simeq 2.1$), corresponding to the dilute regime in which we worked in the previous sections. The decrease in looping exponent $c$ means that the logarithmic potential constraining the diffusive motion of the slip-link is shallower, which leads to a greater efficiency of diffusive loop extrusion (Fig.~\ref{fig8}).
	
Interestingly, as shown in Fig.~\ref{figMSD}C, in our BD simulations, the effective $c$ exponent (measured for loop size between $10$ and $100$~kbp) is not constant, but, for small or intermediate confinement, it becomes progressively smaller over time. This means that as time proceeds the configuration of the polymer slowly evolves. We interpret this slow variation as due to the fact that the motion of the slip-links affects the polymer structure and the looping probability. For the smallest value of $R$, instead, $c$ is approximatively constant, consistent with the polymer being in a long-lived metastable fractal globule which is virtually unaffected by the diffusing slip-links. It would be of interest to repeat these calculations when multiple slip-links bind and unbind to long chromatin fibres, and to see whether the change in $c$ can cause ageing effects, expected for systems with glassy dynamics such as highly confined chromosomes~\cite{Kang2015}, as well as whether it affects the behaviour of the survival probability over time.

As expected, the 1D model does not reproduce the non-monotonic behaviour observed in the 3D BD simulations because there is no change in the friction, or diffusion coefficient, of the 1D random walk as we vary $c$.

We remark that while the set up we have chosen in this section may mimic generic confinement of DNA and chromatin within the nucleus, it also leads to polymer conformations that are statistically similar to those assumed by collapsed polymers, and that have been associated to gene poor or ``heterochromatic'' regions~\cite{Michieletto2016prx,Boettiger2016}. We thus argue that the results of this section suggest that \emph{locally} collapsed chromatin conformations, such as gene poor regions, should display enhanced loop formation kinetics due to diffusive loop extrusion. It should be noted however that these effects may be mitigated by increased microscopic friction.

\section*{Discussion and Conclusions}

In summary, here we have used computer simulations to study the dynamics of diffusive loop extrusion by means of molecular slip-links on chromatin fibres of different flexibility and compaction.

Flexibility is potentially an important parameter which can vary along mammalian chromosomes. The common view is that active regions containing promoters, enhancers and transcribed regions are associated with open chromatin, which is more flexible with respect to that of inactive regions~\cite{Cook2009,Gilbert2004}. Recent microscopy work {\it in vivo} has also shown that the local thickness of the chromatin fibre and its density varies throughout the nucleus~\cite{Ou2017}, and these changes are likely to be associated to a change in flexibility and chromosome folding~\cite{Boettiger2016}.

We have found that the diffusive motion of slip-links, similar to cohesin, is strongly affected by the flexibility and folding of the underlying chromatin fibre. In particular, we have quantified the effect on diffusive loop extrusion -- i.e., the creation of large chromatin loops via diffusive sliding. Whilst such chromatin loops would grow or shrink in the absence of other interactions, assuming that CTCF binds to cohesin in a directionality-dependent manner is sufficient to stabilise these loops, thereby rendering diffusive loop extrusion an appealing model to explain the formation of convergent CTCF loops in mammalian genomes. We have found here that diffusive loop extrusion is substantially faster and more efficient on stiff chromatin, which may be associated with gene poor chromatin~\cite{Sexton2012}. At the same time, more compact polymer conformations and stronger confinement also display enhanced diffusive loop formation kinetics due to a weaker entropic penalty associated with the creation of a loop of a certain size. Our results therefore suggest that cohesin may be an important player to organise inert chromatin regions, where other chromatin bridges are depleted, and in regions of the genome which are locally more compact. 

While the kinetics of loading and unloading are not expected to greatly impact our findings, the competition or cooperations of several cohesin-mediated loops can counteract or enhance the generic principles identified in this work~\cite{Brackley2017prl}.
Specifically, we previously showed that adjacent loops compete with each other for space, while nested loops work cooperatively to enhance the growth of the outermost one. A feature not accounted for in any extrusion model to date is that after DNA has been replicated in S phase, cohesin is responsible for intermolecular links between sister chromatids. This may have a profound impact on loop formation and needs to be considered in future.

Besides being relevant to our understanding of the fundamental mechanisms underlying chromatin looping and 3D chromosome organisation, our results could potentially be tested in single-molecule set-ups with reconstituted chromatin fibres. We also hope they will be of use in designing more sophisticated models of chromatin folding, addressing for instance the lack of steric hindrance and coupling to underlying polymer conformation in current models of active loop extrusion and to understand the interplay between cohesin and other transcription factors (see, e.g.,~\cite{Barbieri2012,Brackley2016nar,Pereira2018}).  Finally, it would be interesting to address the contributions of the proposed enthalpic and entropic effects on the topological regulation of the genome mediated by slip-link-like proteins such as cohesins and condensins~\cite{Orlandini2019}. 

{\it Author contributions:} A.~B., C.~A.~B., J.~J., D.~Michieletto, and D.~Marenduzzo designed the research, performed the research, analyzed the data, and wrote the article.

{\it Acknowledgements:}  This work was supported by ERC (CoG 648050, THREEDCELLPHYSICS).

%%%END OF MAIN TEXT%%%


\balance


%%%REFERENCES%%%
\bibliography{library_new}
\bibliographystyle{rsc} %the RSC's .bst file

\end{document}